\documentclass{iauc}
\usepackage{graphicx}
\usepackage{multicol}
\pubyear{2005}
\volume{199}
\pagerange{1--}
\jname{Probing Galaxies through Quasar Absorption Lines}
\editors{P. R. Williams, C. Shu, and B. M\'{e}nard, eds.}

\newcommand{\lya}{Ly$\alpha$}
\newcommand{\HI}{H\,{\sc i}}
\newcommand{\nhi}{\mbox{$N_{\rm HI}$}}
\newcommand{\Lya}{\mbox{Ly$\alpha$}}
\newcommand{\msun}{\mbox{$M_\odot$}}
\newcommand{\mhi}{\mbox{$M_{\rm HI}$}}
\newcommand{\kms}{\mbox{km s$^{-1}$}}
\newcommand{\cm}{cm$^{-2}$}

\newcommand{\degree}{\hbox{$^\circ$}}

\title[Absorber-galaxy Cross Correlation] 
{Correlation of Low z Lyman-$\alpha$ Absorbers with
HI-selected Galaxies}

\author[Ryan-Weber]   
{Emma Ryan-Weber$^1$}

\affiliation{$^1$Institute of Astronomy, University of Cambridge, UK\break 
email: eryan@ast.cam.ac.uk\\[\affilskip]
}

\begin{document}

\maketitle

\begin{abstract}
  In this work, observational evidence for the connection between low
  column density Lyman-$\alpha$ absorbers and large-scale structure
  traced by gas-rich galaxies is investigated. The \HI\ Parkes All Sky
  Survey (HIPASS) galaxy catalogue is cross-correlated with known low
  redshift, low column density (\nhi$<$$10^{15}$ \cm) Lyman-$\alpha$
  absorbers from the literature. The absorber-galaxy cross-correlation
  function shows that on scales from 1$-$10 h$^{-1}$ Mpc, absorbers
  are imbedded in halos with masses similar to that of galaxy groups.


\end{abstract}

\firstsection 
\section{Introduction}

Hydrodynamic simulations of galaxies and the Intergalactic Medium
(IGM) predict a filamentary structure of the Universe where \HI\ in
the IGM clusters around galaxies (e.g. Dav\'e et
al. 1999). Observations of \HI\ in the IGM via low column density
\lya\ absorption find absorbers in a variety of environments including
large-scale filaments (e.g. Penton et al. 2002, Rosenberg et
al. 2003), galaxy groups (e.g. Bowen et al. 2002), and even voids
(e.g. Stocke et al. 1995). The cross-correction function of \lya\
absorbers and galaxies can be used to measure the extent to which the
two populations of objects, absorbers and galaxies, are associated. In
the Press--Schechter (PS, Press \& Schechter 1974) formalism, the ratio
of the bias of the two populations of objects is equal to the
amplitude ratio of the cross-correlation to the auto-correlation
function (e.g. Mo \& White 2002). Therefore, by (i) knowing the
characteristic total (dark plus baryonic) halo mass (and its relative
bias) of a galaxy population, (ii) measuring the galaxy auto
correlation function, and (iii) measuring the absorber-galaxy
cross-correlation function, the mass of the halos in which the
absorbers are imbedded can be inferred. This method has been
successfully used by \cite{Bouche04}. They cross-correlated MgII
absorbers with Luminous Red Galaxies (LRGs) to show that the MgII
absorbers are imbedded in halos with masses
$\sim2-8\times10^{11}$\msun, consistent with the fact that MgII
absorbers arise in Damped \lya\ (DLA) systems, which are expected to
have total halo masses of that order.

Moving from DLAs to lower column density systems, the minihalo model
(e.g. Mo \& Morris 1994) predicts a weaker absorber-galaxy
cross-correlation function for absorbers associated with the
lower-mass minihalos. The minihalo model has been criticised because
too many halos per unit redshift would be required to account for the
observed density of absorption lines, since the minihalos have small
spatial cross sections. Observational evidence lends more support to
scenarios where absorbers are imbedded in large-scale filaments and
galaxy groups. Furthermore, the PS formalism does not count objects
that have merged into larger collapsed objects (Mo \& Morris 1994).

Perhaps the PS formalism can still work for low column density
absorbers, if instead they are correlated with very large structures.
Here an alternative suggestion is made that if \lya\ absorbers are
imbedded in large scale structure, then their expected clustering
properties should be consistent with large-scale filaments and
galaxy groups. That is, within the PS formalism, a stronger
cross-correlation function amplitude is expected for \lya\ absorbers
on large spatial scales.

Many studies have established a positive association between absorbers
and galaxies (Morris et al. 1993, Stocke et al. 1995, Tripp et al.
1998, Impey et al. 1999, Penton et al. 2002 \& 2004, Bowen et al.
2002), but maintain that a one-to-one correspondence between absorbers
and individual galaxies does not exist in many cases (e.g.  Cot\'e et
al. 2005). To calculate a cross-correlation function, a reasonable
number of absorbers overlapping with a galaxy survey sufficient in
projected distance is needed.

\cite{Morris93} is the only published $z$$=$0 absorber-galaxy
cross-correlation function to-date, which used 17 absorbers along the
3C 273 line-of-sight. They found that absorbers are not distributed at
random with respect to galaxies, but the absorber-galaxy correlation
is weaker than the galaxy auto correlation on scales from 1 $-$ 10
h$_{80}^{-1}$ Mpc. The absorber-galaxy cross-correlation function is
instrumental in establishing whether absorbers are associated with
individual mini-halos or large scale structure. Here, we calculate the
cross-correlation function with many more absorbers.

\firstsection
\section{Cross Correlation Method and Results}

The HI Parkes All Sky Survey (HIPASS) galaxy catalogue (Meyer et
al. 2004, Wong et al. 2005) is cross-correlated with known low
redshift, low column density (mostly \nhi $<10^{15}$ \cm) \Lya\
absorbers from the literature (Impey et al. 1999, Bowen et al. 2002,
Penton et al. 2002 \& 2004, Tripp et al. 2002, Rosenberg et
al. 2003). HIPASS was used as its all sky coverage enables a uniform
statistical comparison with the absorbers, and it identifies gas-rich
dwarf and low surface brightness galaxies, often missed by
magnitude-limited optical surveys. Low redshift \Lya\ absorption
systems were chosen from the literature to overlap with HICAT, i.e.
$\delta$$<$+25\degree\ and heliocentric velocity less than 12~700
\kms, resulting in 141 absorbers from 36 sight-lines. There are 5~340
galaxies in the combined HIPASS catalogues.

The cross-correlation function, $\xi(\sigma,\pi)$ is calculated from
the Davis \& Peebles (1983) estimator,
$\xi(\sigma,\pi)=\frac{DD(\sigma,\pi)}{DR(\sigma,\pi)}\frac{n_R}{n_D}-1$,
where $DD(\sigma,\pi)$ is the number of data galaxy$-$data absorber
pairs with projected separation, $\sigma$, and radial separation,
$\pi$, and $DR(\sigma,\pi)$ is the number of data galaxy$-$random
absorber pairs. The function is normalised by the number of random
absorbers, $n_R$, and data absorbers, $n_D$. The projected correlation
function, $\Xi(\sigma)/\sigma$, is then calculated by integrating
$\xi(\sigma,\pi)$ in the $\pi$ direction, a power-law is then fit to
give the real-space correlation function co-efficients, $r_0$ and
$\gamma$. To statisfy the PS formalism condition that the slopes of
the cross- and auto-correlation function, $\gamma_{ag}$ and
$\gamma_{gg}$, be equal, the co-efficients, $r_{0,gg}$ and
$\gamma_{gg}$, are calculated first, then $r_{0,ag}$ is determined by
fixing $\gamma_{ag}$=$\gamma_{gg}$ and using a Levenberg-Marquardt
non-linear least squares fit. The errors are calculated using
jackknife resampling.

\begin{figure}
\centering 
\includegraphics[height=7cm]{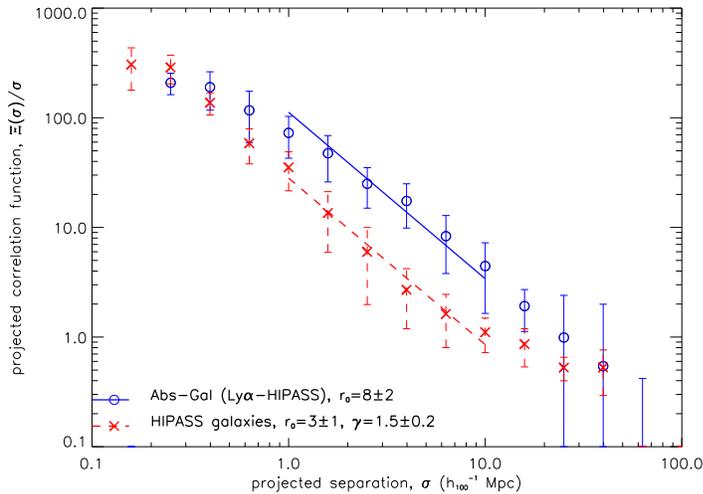} 
\caption{The absorber-galaxy cross-correlation function (circles \&
solid line) and the galaxy auto-correlation function (crosses \&
dashed line). See text for power-law fit description. \label{fig:proj}}
\end{figure}

The auto- and cross-correlation functions are given in
Figure~\ref{fig:proj}. Southern HIPASS has already been used to show
that \HI\ selected galaxies are more weakly clustered than optically
selected galaxies (Meyer, 2003). The results here show that the
absorber-galaxy cross-correlation is stronger than the HIPASS galaxy
auto-correlation for the spatial range 1$-$10 h$^{-1}$ Mpc. The
median \HI\ mass of galaxies contributing to pairs in the range
1$\leq$$\sigma$$\leq$10 h$^{-1}$ Mpc is log(\mhi/\msun)=8.1 h$^{-2}$,
which corresponds to a halo mass of log(M/\msun)=10.7 h$^{-1}$ (Mo et
al., this volume). The ratio of the cross- to auto-correlation function,
$\left(\frac{r_{0,ag}}{r_{0,gg}}\right)^{\gamma_{gg}}$, together with
known halo biases at $z$$=$0 (Mo \& White 2002, Bouch\'e, priv. comm.)
indicates a value of log(M/\msun)$\sim$14 h$^{-1}$ for the mass of
halos in which \Lya\ absorbers are imbedded. This value agrees well
with the median dynamical mass of galaxy groups in HIPASS,
log(M/\msun)$\sim$13.8 h$^{-1}$ (Stevens, 2005).

\firstsection
\section{Summary}

The results show that the absorber-galaxy cross-correlation is
stronger than the galaxy auto-correlation, this is opposite to what is
seen by Morris et al. (1993, based on 17 absorbers). In the context of
the PS formalism, the results suggests that, on scales from 1$-$10
h$^{-1}$ Mpc, \Lya\ absorbers are associated with dark matter halos
with masses of log(M/\msun)$\sim$14 -- similar to that of galaxy
groups. The flattening of $\Xi(\sigma)/\sigma$ at low $\sigma$
suggests that absorbers tend to avoid the densest regions of the local
Universe. Although observing individual systems may reveal an
association of a \lya\ absorber with a particular structure, be it a
galaxy group, filament, void or galaxy halo, the statistical evidence
presented here suggests that galaxy groups could be the dominant
environment of low column density \lya\ absorbers at $z$$=$0. \\

I wish to acknowledge Lister Staveley-Smith \& Rachel Wesbter who
encouraged an early version of this work, and to Nicolas Bouch\'e 
for providing $z$$=$0 bias values.

\begin{multicols}{2}

\end{multicols}

\begin{thebibliography}{}

\bibitem[Bouch\'e et al. (2004)]{Bouche04}
  {Bouch\'e, N. et al.} 2004, MNRAS, 354, L25

\bibitem[]{}Bowen, D.V. et al. 2002, ApJ, 580, 169

\bibitem[]{}Cot\'e, S. et al. 2005, ApJ, 618, 178

\bibitem[{{Dav{\'e}} {et~al.} (1999)}]{Dave99}
{Dav{\' e}}, R. et al. 1999, ApJ,
  511, 521

\bibitem[]{}Davis, M. \& Peebles, P.J.E. 1983, ApJ, 267, 465

\bibitem[{{Impey} {et~al.} (1999)}]{Impey99}
{Impey}, C.~D. et al. 1999, ApJ, 524, 536

\bibitem[Meyer (2003)]{Meyer03}
  {Meyer, M.J.} 2003, PhD Thesis, U. Melbourne

\bibitem[Meyer et al. (2004)]{Meyer04}
  {Meyer, M.J. et al.} 2004, MNRAS, 350 1195

\bibitem[Mo \& Morris (1994)]{Mo94}
  {Mo, H.J, Morris, S.L.} 1994, MNRAS, 269, 52

\bibitem[Mo \& White (2002)]{Mo02}
  {Mo, H.J, White, S.D.M.} 2002, MNRAS, 336, 112 

\bibitem[{{Morris} {et~al.} (1993)}]{Morris93}
{Morris}, S.~L. et al. 1993, ApJ, 419, 524

\bibitem[{{Penton} {et~al.} (2002)}]{Penton02} {Penton}, S.~V.
et al. 2002, ApJ, 565, 720

\bibitem[{{Penton} {et~al.} (2004)}]{Penton04} {Penton}, S.~V.
et al. 2004, ApJ, 152, 29

\bibitem[]{}Press, W.H. \& Schechter, P. 1974 ApJ, 187, 425

\bibitem[]{}Rosenberg, J.L. et al. 2003, ApJ, 591, 677

\bibitem[]{}Stevens, J.B. 2005, PhD Thesis, U. Melbourne

\bibitem[{{Stocke} {et~al.} (1995)}]{Stocke95}
{Stocke}, J.~T. et al. 1995, ApJ, 451, 24

\bibitem[{{Tripp} {et~al.} (1998)}]{Tripp98} {Tripp}, T.~M. et
al. 1998, ApJ, 508, 200

\bibitem[]{Tripp02} {Tripp}, T.~M. et al. 2002, ApJ, 575 697

\bibitem[]{}Wong, O.I. et al. 2005, in preparation.


\end{thebibliography}
\end{document}